\documentclass{llncs}
\usepackage{latexsym}
\usepackage{amsmath}
\usepackage{amssymb}
\usepackage{amsfonts}
\usepackage{epsfig}
\usepackage{color}
\usepackage{times}
\usepackage{graphics}
\usepackage{comment} 
\usepackage{verbatim}

\usepackage[utf8]{inputenc}
\usepackage[T1]{fontenc}

\usepackage{pdflscape}

\usepackage{graphicx}

\usepackage{pst-pdf}

\usepackage[margin=1.6in]{geometry}
\usepackage[]{algorithm2e}

\newcommand{\locations}{\ensuremath{V}}
\newcommand{\aloc}{\ensuremath{loc}}

\newcommand{\pathd}{\ensuremath{P}}

\newcommand{\vehicle}{\ensuremath{veh}}
\newcommand{\ndriver}{\ensuremath{k}}
\newcommand{\orig}{\ensuremath{orig}}
\newcommand{\origin}{\ensuremath{orig}}
\newcommand{\dest}{\ensuremath{dest}}
\newcommand{\tdep}{\ensuremath{dep}}
\newcommand{\tarr}{\ensuremath{arr}}
\newcommand{\mloadd}{\ensuremath{\ell oad}}
\newcommand{\mpathd}{\ensuremath{path}}
\newcommand{\ntourd}{\ensuremath{n}}
\newcommand{\taskset}{\ensuremath{\mathcal{T}}}

\newcommand{\adur}{\ensuremath{dur}}

\newcommand{\abs}[1]{\ensuremath{\lvert #1 \rvert}}

\newcommand{\NN}{\ensuremath{\mathbb{N}}}
\newcommand{\ZZ}{\ensuremath{\mathbb{Z}}}

\newcommand{\tourd}{\ensuremath{\Gamma}}
\newcommand{\action}{\ensuremath{a}}

\newcommand{\move}{\ensuremath{m}}
\newcommand{\dist}{\ensuremath{d}}

\newcommand{\capV}{\ensuremath{\text{Cap}}}

\newcommand{\circuit}{\ensuremath{C}}
\newcommand{\subline}{\ensuremath{L}}

\newcommand{\mods}[1]{{\color{blue}#1}}

\newcommand{\acnum}{\ensuremath{\Delta z}}
\newcommand{\loadd}{\ensuremath{z_m}}
\begin{document}

\begin{center}
\begin{LARGE}
\textbf{Fleet management for autonomous vehicles\\} 
\end{LARGE}
\end{center}

\begin{center}
Sahar Bsaybes, Alain Quilliot and Annegret K.~Wagler\\
Universit\'e Blaise Pascal (Clermont-Ferrand II)\\
   Laboratoire d'Informatique, de Mod\'elisation et d'Optimisation des Syst\`emes \\
   (LIMOS, UMR 6158 CNRS)\\
   BP 10125, 63173 Aubi\`ere Cedex, France\\

\end{center}

\begin{abstract}
The VIPAFLEET project consists in developing models and algorithms for managing a fleet of Individual Public Autonomous Vehicles (VIPA). Hereby, we consider a fleet of cars distributed at 
specified stations in an industrial area to supply internal transportation, where the cars can be used in different modes of circulation (tram mode, elevator mode, taxi mode). 
One goal is to develop and implement suitable algorithms for each mode in order to satisfy all the requests 
under an economic point of view 
by minimizing the total tour length or the makespan. 
The innovative idea and challenge of the project is 
to develop and install a dynamic fleet management system that allows the operator to switch between the different modes within the different periods of the day according to the dynamic transportation demands of the users.
We model the underlying online transportation system and propose an according fleet management framework, 
to handle modes, demands and commands.
We propose for each mode appropriate online algorithms and evaluate their performance. 
\end{abstract}

\section{Introduction}

The project VIPAFLEET aims at contributing to sustainable mobility through the development of innovative urban mobility solutions by means of fleets of Individual Public Autonomous Vehicles (VIPA) allowing passenger
transport in 
closed sites like industrial areas, medical complexes, campuses, business centers, big parkings, airports and train stations. 
A VIPA is an ``autonomous vehicle'' that does not require a driver nor an infrastructure to operate, which reflects the innovative property of the project from the technical side \cite{easymile,viameca}. 
A fleet of VIPAs shall be used in an industrial site 
to transport employees and visitors e.g. between parkings,
buildings and from or to the restaurant at lunch breaks. 
The fleet is distributed at specified stations 
to supply internal transportation, and a VIPA can operate in three different transportation modes:
\begin{itemize}
\item \emph{Tram mode:} VIPAs continuously run on predefined lines or cycles in a predefined direction and stop at a station if requested to let users enter or leave. 
\item \emph{Elevator mode:} VIPAs run on predefined lines 
and react on customer requests by moving to a station to let users enter or leave, thereby changing their driving direction 
if needed. 
\item \emph{Taxi mode:} users book their transport requests (from any start to any destination station within the network with a start and an arrival time) in advance or in real time. 
\end{itemize}
In practice, the different modes lead to different online transportation problems. 

The classical Traveling Salesman Problem (TSP) is a well-known NP-hard problem \cite{papadimitriou1977euclidean}, where a set of cities has to be visited in a single tour, starting and ending in a specified city, with the objective of minimizing the total tour length. This is one of the most studied problems in combinatorial optimization with many variations, see e.g. \cite{gutin2006traveling,lawler1985traveling}. 
In the online version of TSP, the cities are requested to be visited over time \cite{ausiello2004algorithms,IJC:BMK-2001} and a server 
has to decide in which order to serve them, without knowing
the entire sequence of requests in advance. Some cities may not be visited at all or a city may be requested to be visited several times. 

The TSP can be generalized to the $k$-Server Problem where $k$ servers are available to visit cities which includes to partition the requests appropriately and to plan tours for all $k$ servers \cite{bansal2015polylogarithmic,chrobak1991new,manasse1988competitive}. 
In the online version of the $k$-Server Problem, the requests 
are again released over time. 

This can be generalized further to the Pickup-and-Delivery Problem (PDP) where a fleet of servers shall transport goods or persons from a certain origin to a certain destination.  
If persons have to be transported, we usually speak about a Dial-a-Ride Problem. 
Many variations 
are studied 
including the Dial-a-Ride Problem with time windows \cite{deleplanque2013transfers,fabri2007online}. 
In the online case \cite{ascheuer2000online,berbeglia2010dynamic,cordeau2007dial} transportation requests 
are again released over time.

In all mentioned online transportation problems different objective functions can be considered. 
We focus on the economic aspect where the objective is to minimize costs and/or to maximize the profit e.g. by minimizing the total tour length or the makespan.

In a VIPAFLEET system, users can either call a VIPA directly from a station with the help of a call-box 
or can book their request in advance or in real time by mobile or web applications. 
Since the  customer requests are released over time, the classical transportation problems do not reflect the real situation of this project. Therefore we are interested in their online versions. 
Our aim is to develop and install a Dynamic Fleet Management System that allows the operator to switch between different network designs, transportation modes and according online algorithms within the different periods of the day in order to react to changing demands 
evolving during the day, with the objective to satisfy all demands in a best possible way.
For that we model the underlying online transportation system and propose an according fleet management framework, 
to handle modes, demands and commands (see Section 2). Furthermore, in Section 3 we
\begin{itemize}
\item develop suitable algorithms for each combination of mode and subnetwork;
 \item analyze best-case and worst-case behavior of these algorithms for different demands w.r.t different objective functions;
 \item cluster the demands 
into subproblems in such a way that, for each subproblem, a suitable subnetwork and a suitable algorithm can be proposed leading to a globally good solution (transportation schedule).
\end{itemize}
Note that the latter is the innovative idea and challenging problem in the project: in many existing approaches, clustering in subproblems has been proposed (e.g. in \cite{berbeglia2007static}), 
but the same solution technique is applied to each subproblem. Here, we intend the subproblems to be treated by different techniques to improve the global quality of the solution.

\section{Problem description and model}
We embed the VIPAFLEET management problem in the framework of a metric task system.
We encode the closed site  where the VIPAFLEET system is running as a \emph{metric space} $M =(V,d)$ induced by a connected network $G = (V,E)$, where 
the nodes correspond to stations, arcs to their physical links in the closed site, and the distance $d$ between two nodes $v_i,v_j \in V$ is the length of a shortest path from $v_i$ to $v_j$. 
In $V$, we have a distinguished origin $v_o \in V$, the depot of the system where all VIPAs are parked when the system is not running, i.e., outside a certain time horizon $[0,T]$.

A Dynamic Fleet Management System shall allow the operator to switch between different transportation modes within the different periods of the day in order to react to changing customer demands 
evolving during the day. 
For that, for a certain period $[t,t']\subseteq[0,T]$, we define a 
metric subspace $M' =(V',d')$ induced by a subnetwork $G' = (V',E')$ of $G$, 
where a subset of nodes and arcs of the network is active (i.e. where the VIPAs have the right to perform a move on this arc or pass by this node during $[t,t']$).  
An operator has to decide when and how to move the VIPAs in the subnetworks, and to assign customer requests to VIPAs.

Any customer request $r_j$ is defined as a 6-tuple $r_j=(t_j,x_j,y_j,p_j,q_j,z_j)$ where  
\begin{itemize}
\item $t_j \in [0,T]$ is the release date, 
\item $x_j \in V$ is the origin node, 
\item $y_j \in V$ is the destination node, 
\item $p_j \in [0,T]$ is the earliest start time, 
\item $q_j \in [0,T]$ is the latest possible arrival time, 
\item $z_j$ specifies the number of passengers. 
\end{itemize}
Note that a request $r_j$ may have missing information according to the mode wherein a VIPA is operating and to the source of the request. If call-boxes only allow to call a VIPA to pick the user from the station, but not to indicate immediately its destination, then each customer request is split into two: a pickup-request coming from a call-box, and a delivery-request coming from a VIPA after the user has entered.
Accordingly, different types of requests can be distinguished: 
\begin{itemize}
\item \emph{pickup-request:} a request coming from a simple call-box specifying release date and origin $r_j^p=(t_j,x_j,null,null,null,null)$, or simply $r_j^p=(t_j,x_j)$ (only $t_j$ and $x_j$ are known).
\item \emph{delivery-request:} a request coming from a VIPA specifying release date and destination $r_j^d=(t_j,null,y_j,null,null,null)$, or simply $r_j^d=(t_j,y_j)$ (only $t_j$, $y_j$ are known). 
\item \emph{pdp-request:} a request coming from an evolved call-box specifying release date, origin, destination and load, $r_j^{pd}=(t_j,x_j,y_j,null,null,z_j)$, or simply $r_j^{pd}=(t_j,x_j,y_j,z_j)$ ($t_j$, $x_j$, $y_j$ and $z_j$ are known).
\item \emph{full-request:} a request coming from 
a web application $r_j^f=(t_j,x_j,y_j,p_j,q_j,z_j)$, 
where all the parameters $t_j$, $x_j$ , $y_j$, $p_j$, $q_j$, and $z_j$ are known.
\end{itemize}
Note that in the case of delivery-requests (resp. pdp-requests) only destinations $y_j$ can be specified that lie in the same subnetwork where the VIPA is operating (resp. the call-box is installed)\footnote{If the final destination of the customer is outside the subnetwork, he has to change subnetworks and to release further requests.}.  

In particular, the type of the call-boxes has an impact on the request type and therefore on the online transportation problem behind.
\begin{itemize}
\item Having pickup-requests from simple call-boxes and delivery-requests from VIPAs leads to an \textit{Online Traveling Salesman Problem} (one server (the VIPA) has to visit stations in $V$) or to an \textit{Online $k$-Server Problem} (if $k$ servers (VIPAS) are available to visit stations in $V$).
\item Having pdp-requests from evolved call-boxes leads to an \textit{Online Dial-a-Ride Problem} (VIPAs have to transport users from an origin station to a destination in $V$). 
\item Having full-requests from a web application leads to an \textit{Online Dial-a-Ride Problem with Time Windows} (VIPAs have to transport users from an origin station to a destination in $V$ within a certain time window). 
\end{itemize}
For that, the operator monitors the evolution of the requests and the movement of VIPAs over time and creates tasks to move the VIPAS to go to some station and pick up, transport and deliver users.
More precisely, a \emph{task} can be defined in three different ways according to the type of requests that we are dealing with and consequently to the transportation problems behind:
\begin{itemize}
\item \emph{Pickup-task $\tau_j^p = (t_j,x_j,z_j,G^\prime)$:} a task created by the operator in order to serve a pickup-request $r_j^p=(t_j,x_j)$; this task is sent at time $t_j$ to a VIPA operating in the subnetwork $G^\prime$ containing station $x_j$, indicating that $z_j$ passengers have to be picked up (we have {$z_j=1$} since no information is provided).
\item \emph{Delivery-task $\tau_j^d = (t_j,y_j,z_j,G^\prime)$:} a task created by the operator in order to serve a delivery-request $r_j^d=(t_j,y_j)$; this task is sent at time $t_j$ to the VIPA sending the delivery-request $r_j^d$ 
indicating that $z_j$ passengers have to be delivered ($z_j=1$ since no information is provided). 
\item \emph{pdp-task $\tau_j^{pd} = (t_j,x_j,y_j,z_j,G^\prime)$:} a task created by the operator in oder to serve a pdp-request $r_j^{pd}=(t_j,x_j,y_j,z_j)$; this task is sent at time $t_j$ to a VIPA operating in the subnetwork $G^\prime$ containing stations $x_j$ and $y_j$, indicating that $z_j$ passengers have to be picked up at $x_j$ and delivered at $y_j$.
\item \emph{full-task $\tau_j^f = (t_j,x_j,pick,y_j,drop,z_j,G^\prime)$:} a task created by the operator in order to serve a full-request $r_j^f=(t_j,x_j,y_j,p_j,q_j,z_j)$, this task is sent at time $t_j$ to a VIPA operating in the subnetwork $G^\prime$ containing stations $x_j$ and $y_j$, indicating that $z_j$ passengers have to be picked up 
at station $x_j$ at time $pick$ and delivered at station $v$ at time $drop$, 
where $p_j\leq pick\leq q_j-d(x_j,y_j)$ and $p_j+d(x_j,y_j) \leq drop \leq q_j$. 
\end{itemize}
In order to fulfill the tasks, we let a fleet of VIPAs (one or many, each having a capacity for $\capV$ passengers) circulate in the network inducing the metric space. 
An \emph{action} for a VIPA~$j$ is a $5$-tuple $\action = (j, v, t, z, u)$, where 
$j \in \{ 1, \dotsc, \ndriver \}$ specifies the VIPA $\vehicle(\action)$ performing the action, 
$v \in \locations$ specifies the station $\aloc(\action)$, 
$t \in [0, T ]$ is the time $t(a)$ when the action is performed, 
$z \in \ZZ$ the number of users $\acnum(\action)$ to be picked up (if $z > 0$) or delivered (if $z < 0$)
and $u\in \NN$ the duration of the action $\adur(\action)$, the necessary time to pick up or deliver the users.
Hereby, the capacity of the VIPA must not be exceeded, i.e., we have $\abs{z} \leq \capV$.
We say that an action is \emph{performed} (by a VIPA) if $\abs{z}$ users at~$v$ are picked up (resp. delivered) .
A \emph{move} from one station to another is
$\move = (j, v, t^v, w, t^w, \pathd, \loadd)$, where $j \in \{ 1, \dotsc, \ndriver \}$ specifies the VIPA $\vehicle(\move)$ that has to move from the origin station $\origin(\move) = v \in \locations$ 
starting at time $\tdep(\move) = t^v$ to destination station $\dest(\move) = w \in \locations$ arriving at time $\tarr(\move) = t^w$,
a load of $\mloadd(\move) = \loadd$ users in the VIPA moving along the path $\mpathd(\move) = \pathd$.
Hereby, we require that
\begin{itemize}
 \item\label{def: enum: move: 3} $0 \leq \mloadd(\move^i) \leq \capV$,
 \item\label{def: enum: move: 2} from $\orig(\move) \neq \dest(\move)$ follows $\tarr(\move) = \tdep(\move) + \dist(\orig(\move), \dest(\move))$.
\end{itemize}
A \emph{tour} is a sequence $\tourd = (\move^1, \action^1, \move^2, \action^2, \dotsc, \action^{\ntourd - 1}, \move^\ntourd)$ of moves and actions with
\begin{itemize}
 \item\label{def: enum: tour: 1} $\vehicle(\move^1) = \vehicle(\action^1) = \dotsm = \vehicle(\action^{\ntourd - 1}) = \vehicle(\move^\ntourd)$, 
 \item\label{def: enum: tour: 2} $\dest(\move^i) = \aloc(\action^i) = \orig(\move^{i+1})$,
 \item\label{def: enum: tour: 3} $\tarr(\move^i) = t(\action^i),$
 \item\label{def: enum: tour: 4} $\tdep(\move^{i+1})= t(\action^i) + dur(\action^i)$,
 \item\label{def: enum: tour: 5} $\mloadd(\move^{i+1}) = \mloadd(\move^{i}) + \acnum(\action^i)$,
 \item\label{def: enum: tour: 6} for every move $\move^i$, $\orig(\move^i)$, $\dest(\move^i)$ and $\mpathd(\move^i)$ lie on the same subnetwork $G^\prime$.
\end{itemize}

A subsequence of a tour is called a \emph{subtour}.
Each tour is a sequence of two or more subtours: the initial subtour, from the depot to the origin of the subnetwork where the VIPA will operate, the final
subtour from the origin of the subnetwork back to the depot and the intermediate subtours (if any) from the origin of the subnetwork back to it.

Several tours are composed to a transportation schedule.
A collection of tours $\{\tourd^1, \ldots, \tourd^\ndriver \}$ is a \emph{transportation schedule} $S$ for 
$(M, \taskset)$ if 
\begin{itemize}
 \item\label{def: enum: schedule: 1} every VIPA has exactly one tour,
 \item\label{def: enum: schedule: 2} each request  $r_i \in R$  is served not earlier than the time it is released.
 \item each tour starts and ends in the depot.
\end{itemize}
If each user is transported from its start station to its final destination by only one VIPA, then $S$ is called non-preemptive, otherwise preemptive. 
In particular, if start and destination station of a user do not lie on the same subnetwork $G^\prime$, the user has to change VIPAs on one or more intermediate stations. 
As this typically leads to inconveniences, the design of subnetworks has to be done in such a manner that preemption can be mostly avoided. 

In addition, depending on the policy of the operator of such a system, different side constraints have to be obeyed.
If two or many VIPAs circulate on the same network, the fleet management has to handle e.g. the 
\begin{itemize}
 \item meeting of two vehicles on a station or an arc,
 \item blocking the route of a VIPA by another one waiting at a station (if two VIPAs are not allowed to enter the same node or arc at the same time),
\end{itemize}
and has to take into account 
\begin{itemize}
 \item the events of breakdown or discharge of a vehicle,
 \item technical problems with the server, the data base or the communication network between the stations, VIPAs and the central server. 
\end{itemize}

Our goal is to construct transportation schedules $S$ for the VIPAs respecting all the above constraints w.r.t minimizing one of the following objective functions. 
\begin{itemize}
 \item \emph{Total tour length}: the length of a tour corresponds to the distance traveled by the VIPA, the total tour length of $S$ is the sum of the lengths of its tours without considering the waiting time of the VIPAs.
 \item \emph{Makespan}: the time when all VIPAs returned to the depot $v_0$ after all requests are served.
\end{itemize}

This will be addressed by dividing the time horizon $[0,T]$ in different periods according to the volume and kind of the requests, and by providing specific solutions within each period.
The global goal is to provide a feasible transportation schedule over the whole time horizon that satisfies all requests and minimizes one of these objective functions (Dynamic Fleet Management Problem). 
For each period, partition the network $G = (V,E)$ 
into a set of subnetworks $G^\prime = (V^\prime,E^\prime)$.
The aim of this partition is 
to solve some technical side constraints for autonomous vehicles, 
to make use of the different types of requests, and therefore to be able to solve different transportation problems using different algorithms at the same time on the same network, and to gain precision
in solutions by applying a suitable algorithm to a certain subnetwork.
The choice of the design of the network in the industrial site where the VIPAs will operate is dynamic and will change over time according to the technical features and properties.

In a metric 
space we partition the network 
into subnetworks $G^\prime$ that are either unidirected cycles, called \textit{circuits} $G^\prime_c$ or bidirected paths, called \textit{lines} $G^\prime_{\ell}$. This partition is motivated by two of the operation modes for VIPAs: tram mode and elevator mode. 
To operate VIPAs in taxi mode, the whole network or a sparse connected subnetwork will be used. 

\section{Scenarios, algorithms and competitive analysis}

Based on all the above technical features and properties that have an impact on the feasibility of the transportation schedule, we can cluster the requests into subproblems, apply to each subproblem a certain algorithm, and check the results in terms of feasibility and performance. 

\subsection{Combinations of modes and subnetworks for different scenarios}

The choice of the design of the network in the industrial site  where the VIPAs will operate is dynamic and will change over time according
to the technical features and properties. Let us consider four typical scenarios that might happen while operating a fleet in an industrial site, based on some preliminary studies of the transport requests within the site.
 
\paragraph{Morning/evening:} 
The transport requests are between parkings and buildings. 
For this time period we propose the following:
 \begin{itemize}
  \item [$\bullet$] Design a collection of subnetworks (lines and circuits) s.t.
  \begin{itemize}
   \item [-] all buildings and parkings are covered,
   \item [-] each subnetwork contains one parking $p$ and all the buildings where $p$ is the nearest parking (to ensure that for each request, origin (the parking) and destination (a building) lie in the same subnetwork).
  \end{itemize}
  \item [$\bullet$] Depending on the number of employees in the served buildings, assign one VIPA (in elevator mode) to every line and one or several VIPAs (in tram mode) to each circuit.

 \end{itemize}

\paragraph{Lunch time:}
The transport requests are between buildings and the restaurant of the industrial complex. 
For this time period, we propose the following:
 \begin{itemize}
  \item [$\bullet$] Design a collection of lines s.t.
  \begin{itemize}
   \item [-] all buildings are covered,
   \item [-] each line contains the station of the restaurant (to ensure that for each request, to or from the restaurant, origin and destination lie in the same line).
  \end{itemize}
  \item [$\bullet$] Depending on the number of employees in the served buildings, assign one VIPA (in elevator mode) or one or several VIPAs (in tram mode) to the lines.
 \end{itemize}

\paragraph{Emergency case:} 
In the case of a breakdown of the central servers, the database or the communication system, transports between all possible origin/destination  pairs have to be ensured without any decision by the operator. For that we propose
 \begin{itemize}
  \item to use one Hamilton cycle through all the stations as subnetwork and 
  \item to let half of the fleet of VIPAs operate in each direction on the cycle (all in tram mode).
 \end{itemize}

\paragraph{Other periods:} 
There are mainly unspecified requests without common origins or common destinations. 
 The operator can use all VIPAs in his fleet in taxi mode on the complete network or design lines and circuits s.t. all stations are covered and the chosen subnetworks intersect (to ensure transports between all 
 possible origin/destination pairs). E.g., 
this can be done by
 \begin{itemize}
  \item using one Hamilton cycle through all stations where half of the fleet operates (in tram mode) in each direction,
  \item a spanning collection of lines and circuits meeting in a central station where one VIPA (in elevator mode) operates on each line, one or several VIPAs (in tram mode) on each circuit.
 \end{itemize}

\subsection{Online algorithms and their analysis}

Recall that the customer requests are released over time s.t. the studied transport problems have to be considered in their online version. 
A detailed introduction to online optimization can be found e.g. in the book by Borodin and El-Yaniv \cite{borodin2005online}.
It is standard to evaluate the quality of online algorithms with the help of competitive analysis.
This can be viewed as a game between an online algorithm $ALG$ and a malicious adversary who tries to generate a
worst-case request sequence $\sigma$ which maximizes the ratio between the online cost $ALG(\sigma)$ and the optimal cost $OPT(\sigma)$
knowing the entire request sequence $\sigma$ in advance.
 $ALG$ is called \emph{$k$-competitive} if $ALG$ produces for any request sequence $\sigma$ a feasible solution 
 with
\[
 ALG(\sigma)  \leq k \cdot OPT (\sigma)
\]
for some given $k \geq 1$. The competitive ratio of ALG is the infimum over all $k$ such that $ALG$ is $k$-competitive.
We are interested in designing and analyzing online algorithms for each possible operating mode of a VIPA.

\paragraph{Tram mode:}
The tram mode is the most restricted operation mode where VIPAs run on predefined circuits in a predefined direction and stop at stations to let users enter or leave.
 The behavior of the VIPAs 
is even independent of the type of call-boxes used and, thus, the type of generated requests and tasks.
 We consider circuits $\circuit$ with one distinguished node, the origin of the circuit\footnote{Circuits can also be lines where one end is the origin and the VIPA can change direction only in the two ends of the line.}.
 We propose the following simple algorithm for VIPAs operating in tram mode on a circuit $\circuit$:\\

\noindent
SIR (``Stop If Requested'')\\[-6mm] 
\begin{itemize}
  \item each VIPA waits in the origin of $\circuit$; 
as soon as a request is released, a VIPA starts a full subtour in a given direction, thereby it stops at a station when a user requests to enter/leave.
 \end{itemize}
In fact, in tram mode, the possible decisions of the server are either to continue its tour or to wait at its current position for newly released requests.
This can be used by the adversary to ``cheat'' an online algorithm, in order to maximize the ratio between the online and the optimal costs.

We firstly consider the objective of minimizing the total tour length. 
Here, the strategy of the adversary is to force SIR to serve only one request per subtour, whereas the adversary only needs a single subtour to serve all requests.

\begin{example}
\label{exa: comp: SIR: CAPC-totaltourlength}
Consider a circuit $\circuit=(v_1,\dotsc,v_n)$ with origin $v_1$, a unit distance between $v_i$ and $v_{i+1}$ for each $i$, and one unit-speed server
(i.e. a VIPA that travels 1 unit of length in 1 unit of time).
The adversary releases  a sequence $\sigma$ of $\capV \cdot n$ pdp-requests that force SIR to perform one full round (subtour) of length $\abs{C}=n$ for each request, whereas the adversary is able to serve all requests in a single subtour:
\begin{itemize}
 \item $\capV$ requests $r_i=((i-1)\abs{C},v_1,v_2,1)$ for $1 \leq i \leq \capV$
 \item $\capV$ requests $r_i=((i-1)\abs{C},v_2,v_3,1)$ for $\capV+1 \leq i \leq 2\capV$\\
 $\vdots$
 \item $\capV$ requests $r_i=((i-1)\abs{C},v_{n-1},v_n,1)$ for $(n-2)\capV+1 \leq i \leq (n-1)\capV$
 \item $\capV$ requests $r_i=((i-1)\abs{C},v_n,v_1,1)$ for $(n-1)\capV+1 \leq i \leq n\capV$
\end{itemize}
SIR starts its VIPA at time $t=0$ to serve $r_1=(0,v_1,v_2)$ and finishes the first subtour of length $\abs{C}$ without serving any further request.
When the VIPA operated by SIR is back to the origin $v_1$, the second request $r_2=(\abs{C},v_1,v_2)$ is released and SIR starts at $t=\abs{C}=n$ a second subtour of length $\abs{C}$ to serve $r_2$,
without serving any further request in this subtour. This is repeated for each request yielding
\[
  SIR(\sigma)=\capV \cdot \abs{C} \cdot \abs{C}.
\]
The adversary waits at the origin $v_1$ until $t=(\capV -1)\abs{C}$ and serves all $r_1,\dotsc,r_{\capV}$ from $v_1$ to $v_2$. Then he waits until $t=(2\capV -1)\abs{C}$ at $v_1$ and serves all requests 
$r_{\capV+1},\dotsc, r_{2\capV}$ from $v_2$ to $v_3$. This is repeated for all $\capV$ requests from $v_i$ to $v_{i+1}$, yielding 
\[
  OPT(\sigma)=\abs{C}.
\]
The tours performed by SIR and OPT are illustrated in Fig~\ref{fig: SIR-worst-case-ten}.
Therefore, we obtain \[\frac{SIR(\sigma)}{OPT(\sigma)} = \frac{\capV \cdot \abs{C} \cdot \abs{C}}{\abs{C}}= \capV \cdot \abs{C}\]
as a lower bound for the competitive ratio of SIR.
\begin{figure}[ht]
    \centering
    \includegraphics[width=0.99\textwidth]{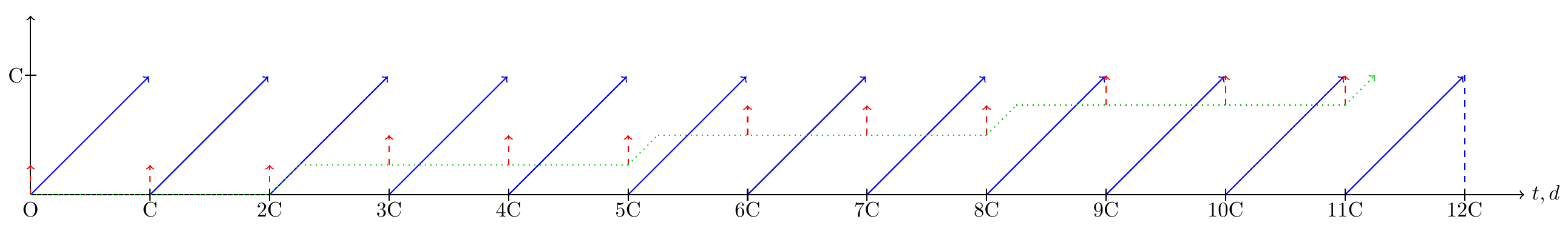}
 \caption{This figure illustrates the tour performed by SIR (in blue) and the adversary (dotted in green) in order to serve the requests (dashed arcs in red) from Example~\ref{exa: comp: SIR: CAPC-totaltourlength} for $\capV = 3, n=4$.}
 \label{fig: SIR-worst-case-ten}
\end{figure}
\end{example}
\vspace{-0.5cm}
The previous example provides the worst case for SIR w.r.t minimizing the total tour length:
\begin{theorem}
SIR is $\capV \cdot \abs{C}$-competitive w.r.t minimizing the total tour length for one or several VIPAs operating in tram mode on a circuit $\circuit$ with length $\abs{C}$.
\end{theorem}

In the special case of the morning scenario, we consider VIPAs operating in tram mode on a circuit $\circuit$ where the parking is the distinguished origin of $\circuit$. 
A sequence $\sigma'$ containing the first $\capV$ requests from the sequence presented in Example~\ref{exa: comp: SIR: CAPC-totaltourlength} shows that $\capV$ is a lower bound on the competitive ratio of SIR. We can show that this is the worst case: 
 
\begin{theorem}
SIR is $\capV$-competitive w.r.t minimizing the total tour length for one or several VIPAs operating in tram mode on a circuit $\circuit$ with length $\abs{C}$ during the morning scenario.
\end{theorem}

Moreover, SIR can be adapted to follow the strategy of the adversary. 
For that, we propose another algorithm for VIPAs operating in tram mode in the morning:\\ 

\noindent
$SIF_M$ (``Start if fully loaded'')\\[-6mm]  
\begin{itemize}
 \item each VIPA waits in the parking until $\capV$ passengers have entered,
 \item it starts a full round (as soon as it is fully loaded) and stops at stations where passengers request to leave.
\end{itemize}

We can ensure optimality of this strategy for the morning:
\begin{theorem}
$SIF_M$ is $1$-competitive w.r.t minimizing the total tour length for one or several VIPAs operating in tram mode on a
circuit during the morning scenario.
\end{theorem}

For the evening scenario, we can also provide an optimal strategy, provided that evolved call-boxes are installed along the circuit (such that the number of waiting customers is known):\\

\noindent
$SIF_E$ (``Start if fully loaded'')\\[-6mm]  
\begin{itemize}
 \item each VIPA waits in the parking until enough requests are released to reach $\capV$,
 \item it starts a full round and stops at stations where passengers request to enter and returns (fully loaded) to the parking.
\end{itemize}

We can ensure optimality of this strategy for the evening:
\begin{theorem}
$SIF_E$ is $1$-competitive w.r.t minimizing the total tour length for one or several VIPAs operating in tram mode on a
circuit during the evening scenario.
\end{theorem}
Concerning the objective of minimizing the makespan, the adversary can cheat SIR by releasing a request on a station where the VIPA operated by SIR shortly left.
\begin{example}
Consider a circuit $\circuit=(v_1,\dotsc, v_n)$ with origin $v_1$ and one unit-speed server.
The adversary releases a sequence $\sigma=(r_1,r_2)$ with only 2 requests
\begin{itemize}
\item[] $r_1=(0,v_1,v_n,1)$,
\item[] $r_2=(\varepsilon,v_1,v_n,1)$.
\end{itemize}
SIR starts its VIPA at time $t=0$ to serve $r_1$. It returns to $v_1$ at time $t=\abs{C}$ and starts a second round to serve $r_2$, yielding 
\[
  SIR(\sigma)=2\abs{C}.
\]
The adversary waits at the origin $v_1$ until $t=\varepsilon$, starts its VIPA with both requests $r_1$ and $r_2$ and is back to $v_1$ at time $t=\abs{C}+\varepsilon$, yielding
\[
  OPT(\sigma)=\abs{C}+\varepsilon.
\]
\end{example}
This gives a lower bound of 2 for the competitive ratio of SIR w.r.t. minimizing the makespan, even during the morning scenario.
The same is true for $SIF_E$ during the evening:
\begin{example}
Consider a circuit $\circuit=(v_1,\dotsc, v_n)$ with origin $v_1$, a unit distance between $v_i$ and $v_{i+1}$ for each $i$, and one unit-speed server.
 The adversary releases a single request
 \begin{itemize}
  \item [] $r= (n-1,v_n,v_1,\capV)$
 \end{itemize}
with $load(r)=\capV$. $SIF_E$ starts its VIPA at time $t=n-1$ to serve $r$, arrives at $t=2(n-1)$ at $v_n$ and is back to $v_1$ at time $t=2n-1$, yielding $SIF_E(\sigma)=2n-1$.
The adversary starts its VIPA at $t=0$, arrives at $v_n$ at $t=n-1$ (when $r$ is released)  and is back to $v_1$ at time $t=n$, yielding $OPT(\sigma)=n$.
\end{example}
This also gives a lower bound of 2 the competitive ratio of $SIF_E$ w.r.t minimizing the makespan.
Just for $SIF_M$, the situation might be better. We conjecture that the following sequence is a worst case for $SIF_M$:
\begin{example}
Consider a circuit $\circuit=(v_1,\dotsc, v_n)$ with origin $v_1$ and one unit-speed server.
 The adversary releases a sequence $\sigma$ with $3$ pdp-requests
\begin{itemize}
\item[] $r_1=(0,v_1,v_n,1)$,
\item[] $r_2=(n,v_1,v_n,\capV-1)$,
\item[] $r_3=(n,v_1,v_n,1)$,
\end{itemize}
announcing that $r_3$ is the last request in $\sigma$.
$SIF_M$ starts its VIPA at time $t=n$ fully loaded to serve $r_1$ and $r_2$, and finishes the first subtour at time $t=2n$.
Because $r_3$ is the last request in $\sigma$, $SIF_M$ starts a second subtour to serve $r_3$ (without being fully loaded) at time $t=2n$ and is back to $v_1$ at time $t=3n$, yielding
$SIF_M(\sigma)=3n$.
The adversary starts its VIPA directly at $t=0$ to serve $r_1$, is back to the origin $v_1$ at time $t=n$ and can immediately serve $r_2$ and $r_3$ together in a second subtour, finishing
 its tour at $t=2n$, yielding $OPT(\sigma)=2n$.

\end{example}
This shows that $3/2$ is a lower bound for the competitive ratio of $SIF_M$ w.r.t minimizing the
makespan in the morning. We conjecture that this bound is tight.

\paragraph{Elevator mode:}
The elevator mode is a less restricted operation mode where one VIPA runs on a predefined line and can change its direction at any station of this line to move towards a requested station.
One end of this line is distinguished as origin $O$ (say, the ``left'' end).

An algorithm MRIN (``Move Right If Necessary'') has been proposed for Online-TSP on a line in \cite {IJC:BMK-2001} 
and analyzed w.r.t. minimizing 
the makespan, giving a competitive ratio of $3/2$.
These results carry over to VIPAs operating in elevator mode on a line where simple call-boxes are installed.
We generalize MRIN further to the situation considering pdp-requests coming from evolved call-boxes, leading to an Online-DARP.
In elevator mode, the server (VIPA) has the choice to continue its tour in the current direction, to wait at its current position or to change its driving direction.
Accordingly, we propose the algorithm MAIN.
\begin{algorithm}
MAIN (``Move Away If Necessary'')\\
\SetKwInOut{Input}{Input}\SetKwInOut{Output}{Output}
\Input{ request sequence $\sigma$, line $\subline$ with origin $O$, $\capV$ }
\Output{tour on $\subline$ to serve all requests in $\sigma$ }
 \emph{current server position $s:=O$}\;
 \emph{set of currently waiting requests (already released requests but not yet served) $\sigma^\prime:=\{r_j\in \sigma: t_j=0\}$}\;
  \While{ $\sigma^\prime \neq \varnothing$}{
  determine the subset $\sigma^\prime_A$ of requests $r_j=(t_j,x_j,y_j)\in \sigma^\prime $ with $s\leq x_j \leq y_j$\;
  \If{ $\sigma^\prime_A \neq \varnothing$}
  { Serve all (or the first $\capV$) requests in $\sigma^\prime_A$ (moving away from $O$ to furthest destination $y_k$ among all $r_j\in \sigma^\prime_A$)\;}
 \Else{
  determine subset $\sigma^\prime_O$ of requests $r_j=(t_j,x_j,y_j)\in \sigma^\prime $ with $x_j > y_j$
  serve all  (or the first $\capV$) requests in $\sigma^\prime_O$ while moving to the origin\;
  }
  update $s$ and $\sigma^\prime$ (remove all served requests, add all newly released requests)\;
}
\end{algorithm}\DecMargin{1em}

The adversary can again ``cheat'' the strategy of MAIN, as the following example shows.


 
\begin{example}
\label{exa: comp: MAIN: comp-makespan}
Consider a line $\subline=(v_0,\dotsc,v_n)$ with origin $v_0$,  a unit distance between $v_i$ and $v_{i+1}$ for each $i$, and one unit-speed server.
The adversary releases a sequence $\sigma=(r_1,r_2)$ with 2 requests
\begin{itemize}
\item[] $r_1=(0,v_0,v_n,1)$,
\item[] $r_2=(\varepsilon,v_0,v_n,1)$.
\end{itemize}
MAIN determines at time $t=0$ the set $\sigma^\prime_A=\{r_1\}$ and serves $r_1$ by moving from $v_0$ to $v_n$. At time $t=n$, we have $s=n$ and $\sigma^\prime_A=\varnothing$, so it moves back to $v_0$, 
arriving at time $t=2n$. Now, $s=v_0$ and $\sigma^\prime_A=\{r_2\}$, so MAIN starts its VIPA again to serve $r_2$ by moving from $v_0$ to $v_n$, and finally returns to $v_0$ at time $t=4n$.
The adversary waits in $v_0$ until $t=\varepsilon$ and serves both requests $r_1$ and $r_2$ by moving from $v_0$ to $v_n$, and returns to $v_0$ at time $t=2n+\varepsilon$.
Therefore we obtain
\[
\frac{MAIN(\sigma)}{OPT(\sigma)} = \frac{4n}{\varepsilon+2n}= 2
\]
as a lower bound for the competitive ratio of MAIN w.r.t. minimizing the makespan, even in the morning scenario.
\end{example}

In a similar way, one can construct a sequence for the evening scenario to show that the competitive ratio of MAIN w.r.t. minimizing the makespan is at least 2.
We conjecture that this bound is tight in general and can ensure it for the morning scenario:
\begin{theorem}\label{th-MAIN-makespan}
MAIN is $2$-competitive for Online-DARP w.r.t minimizing the makespan for one VIPA operating in elevator mode
on a line during the morning scenario. 
\end{theorem}

Concerning the objective of minimizing the total tour length, it turns out that the competitive ratio of MAIN is higher. 
In fact, a similar sequence $\sigma$ as in Example \ref{exa: comp: SIR: CAPC-totaltourlength} for SIR can be used to show that the adversary can force the VIPA operated by MAIN to oscillate between $v_1$ and $v_i$ to serve a single request, for each $i$ with $2 \leq i \leq n$, yielding $MAIN(\sigma)=n(n-1)\capV$, whereas the optimal solution is like in Example \ref{exa: comp: SIR: CAPC-totaltourlength} with $OPT(\sigma)=2n$. 
This shows that the competitive ratio of MAIN w.r.t minimizing the total tour length is at least $\frac{1}{2}(n-1)\capV$ in general. We can show that the situation improves for the special case of the morning scenario:

\begin{theorem}\label{th-MAIN-cap}
MAIN is $\capV$-competitive for Online-DARP w.r.t minimizing the total tour length for one VIPA operating in elevator mode on a line during the morning scenario. 
\end{theorem}

\section{ Concluding Remarks} 

Vehicle routing problems integrating constraints on autonomy are new in the field of operational research but are important for the future mobility.
Autonomous vehicles, which are intended to be used as a fleet in order to provide a transport service, need to be effective also considering to their management.
The future works are to analyze the proposed scenarios and algorithms further, e.g., by
\begin{itemize}
\item providing competitive ratios for the lunch scenario,
\item studying also the quality of service aspect by minimizing the waiting time for customers. 
\end{itemize}
Competitive analysis has been one of the main tools for deriving worst-case bounds on the performance of algorithms but an online algorithm having the best competitive ratio in
theory may reach the worst case more frequently in practice with a certain topology. 
That is the reason why we are not only  interested in the ``worst-case'' but also in the ``best-case''  performance of the algorithms, 
thus we need to determine properties which govern the behavior of each chosen algorithm and define the cases where it can be applied and give the best results in terms of performance.

\bibliographystyle{plain}
\providecommand{\availatURL}[1]{\ignorespaces \footnote{Avail. at URL
  \texttt{#1}}} \providecommand{\NP}{\textsf{NP}}
\providecommand{\bysame}{\leavevmode\hbox to3em{\hrulefill}\thinspace}
\providecommand{\MR}{\relax\ifhmode\unskip\space\fi MR }
\providecommand{\MRhref}[2]{%
  \href{http://www.ams.org/mathscinet-getitem?mr=#1}{#2}
}
\providecommand{\href}[2]{#2}

\bibliography{arxiv-2016}

\end{document}